\documentclass[preprint,showpacs,showkeys,amsmath,amssymb,aps,doublespace]{revtex4}
\usepackage[latin1]{inputenc}
\usepackage{epsfig}
\usepackage{amsmath}
\usepackage{amsfonts}
\usepackage{amssymb,color}
\usepackage{graphicx}
\usepackage{pstricks,pst-grad}

\begin{document}
\title{Conserved directed percolation: exact quasistationary distribution of small systems
and Monte Carlo simulations}
\author{J\'ulio C\'esar Mansur Filho\footnote{e-mail:
jcmansur@fisica.ufmg.br} and Ronald Dickman}

\affiliation{Departamento de F\'{i}sica, Instituto de Ci\^{e}ncias
Exatas, and
National Institute of Science and Technology for Complex Systems,
Universidade Federal de Minas Gerais, CP 702, CEP 30161-970,
Belo Horizonte, Minas Gerais, Brasil.}

\begin{abstract}
We study symmetric sleepy random walkers, a model exhibiting an
absorbing-state phase transition in the conserved directed
percolation (CDP) universality class. Unlike most examples of this
class studied previously, this model possesses a continuously
variable control parameter, facilitating analysis of critical
properties. We study the model using two complementary approaches:
analysis of the numerically exact quasistationary (QS) probability
distribution on rings of up to $22$ sites, and Monte Carlo
simulation of systems of up to $32000$ sites. The resulting
estimates for critical exponents $\beta$, $\beta/\nu_\perp$,
and $z$, and the moment ratio
$m_{211} = \langle \rho^2 \rangle/\langle \rho \rangle^2$
($\rho$ is the activity density), based on finite-size scaling at
the critical point, are in agreement with previous results for the
CDP universality class. We find, however, that the approach to the
QS regime is characterized by a different value of the dynamic
exponent $z$ than found in the QS regime.
\end{abstract}

\maketitle
\section{Introduction}

Over the last several decades, phase transitions between an active and an absorbing
state have attracted great interest in statistical physics and related
fields \cite{Hinrichsen,livrodickman,lubeck,geza}.  More recently, experiments
on such transitions have been performed \cite{takeuchi,bacterias,pine1}.
As in equilibrium, continuous phase transitions to an absorbing state can be grouped into
universality classes \cite{lubeck,geza}. Two
classes that have received much attention are directed percolation
(DP) and conserved directed percolation (CDP), exemplified, respectively, by the
contact process \cite{harris} and the stochastic conserved sandpile (conserved Manna model)
\cite{manna,manna1d}.  While the former class is well characterized, and there is a
clear, consistent picture of the scaling behavior, the critical exponents of CDP have
not been determined to high precision, and there are suggestions of violations of
scaling.  Thus it is of interest to study further examples of this class, and to
apply new methods of analysis to such models.
Absorbing-state transitions have been studied via
mean-field theory, series expansion \cite{livrodickman},
renormalization group \cite{padrao}, perturbation theory
\cite{pertubacao} and numerical simulation.  Recently, an analysis
based on the exact (numerical) determination of the quasistationary
(QS) probability distribution was proposed and applied to models in the DP class \cite{exact sol}.

In this paper we study sleepy random walkers (SRW), a Markov process
defined on a lattice, belonging to the CDP universality class,
using exact (numerical) analysis of the quasistationary
(QS) probability distribution and Monte Carlo simulation.
The former approach furnishes quite accurate predictions
for the contact process; a preliminary application to a model in the
CDP class yielded less encouraging results, due in part to the small
system sizes accessible \cite{exact sol}.  The smaller number of configurations (for a given
lattice size and particle density) in the SRW model allows us to study somewhat larger
systems, leading to improved results in the QS analysis.
We study the model in extensive Monte Carlo simulations as well, in efforts
to better characterize CDP critical behavior.
A closely related model, {\it activated random walkers} (ARW), was introduced in \cite{vladas};
in this case there is no restriction on the number of walkers per site.
In \cite{vladas} the principal emphasis
was on {\it asymmetric} ARW (hopping in one direction only); some
preliminary evidence for CDP-like behavior of the symmetric version
was also reported.

The balance of this paper is organized as follows.  In Sec. II we define the model and
its behavior in mean-field theory.
In Sec. III we describe how exact QS analysis is applied to the model and present
the associated results on critical behavior.  We report our simulation results in
Sec. IV, and
in Sec. V we present a summary of our findings.

\section{Model}

The SRW model is defined on a $d$-dimensional
lattice of $L^d$ sites with periodic boundary conditions. Each site $i$
of the lattice may be in one of three states: empty ($\sigma_i=0$),
occupied by an active particle ($\sigma_i=1$), or by an inactive particle
($\sigma_i=-1$). Multiple occupancy is forbidden. Active particles attempt
to hop, at unit rate, to a nearest-neighbor site.  In a
hopping move the target site is chosen with uniform probability on the
set of nearest neighbors, and the move is accepted if and only if the target
site is vacant. Transitions from $\sigma_i=1$ (active) to
$\sigma_i=-1$ (inactive) occur at a rate of $\lambda$, called
the {\it sleeping rate}, independent of the states of
the other sites. Inactive particles cannot hop. A transition from
$\sigma_i=-1$ to $\sigma_i=1$ occurs when
an active neighbor attempts to jump to site $i$. In this case, the
particle that attempted to hop returns to its original site, but the
particle that was sleeping is activated. Evidently this Markovian
dynamics conserves the number of particles.  Here we consider initial configurations
in which $N$ particles (all active) are distributed randomly amongst the sites,
respecting the prohibition of multiple occupancy.
(In the ARW model \cite{vladas} the number of particles per site is
unrestricted but only an isolated particle can sleep.)

Let $N_{a}$ denote the number of active particles;
any configuration with $N_a = 0$ is absorbing.  Thus we define the
order parameter as $\rho \equiv N_{a}/N$, the fraction of active
particles.  There are two control parameters, the sleeping rate
$\lambda$ and the particle density $\zeta = N/L^d$.
For $\zeta < 1$, the particle number is a nontrivial conserved quantity,
and we expect the model to belong to the CDP universality class.  (For
$N=L$ particle conservation follows trivially from the conservation of
site number, and the model is equivalent to the contact process, belonging to the DP class.)

An advantage of this model is the presence of a {\it continuously variable}
control parameter, $\lambda$.  In the stochastic sandpile \cite{manna,manna1d},
the control parameter, $\zeta$, cannot only be varied in increments of
$1/L^d$, which tends to complicate the determination of critical properties.
(A given value of $\zeta$ is accessible only for a restricted set of system sizes.)
We shall therefore fix the particle density and vary $\lambda$ to locate the
critical point.  Since the QS distribution analysis depends on applying
finite-size scaling analysis, we use a $\zeta=1/2$, which is accessible in all systems
with $L$ even.

Mean field (MF) analysis yields the following equation of motion
for the fraction of active particles:

\begin{equation}\label{eqmp}
    \frac{d \rho}{dt}=(\zeta -\lambda) \rho - \zeta \rho^2,
\end{equation}

\noindent which is analogous to the MF equation for the contact process (CP) \cite{livrodickman}
if we identify $\zeta$ and $\lambda$ with the creation and annihilation rates,
respectively, in the CP.  One sees immediately that at this level
of approximation, an active stationary state exists only for $\lambda < \lambda_c = \zeta$,
in which case the stationary order parameter is $\overline{\rho} = \zeta - \lambda$.
Although the MF analysis is certainly not reliable in detail, it is reasonable to
expect that the model exhibits a continuous phase transition, and that $\lambda_c$
is an increasing function of $\zeta$.

\section{Quasistationary analysis}\label{secQS}

\subsection{Quasistationary probability distribution}

In \cite{exact sol}, one of us proposed a method for studying absorbing-state
phase transitions based on numerical determination of the
quasistationary probability distribution, that is, the asymptotic
distribution, conditioned on survival.
With the essentially exact QS properties in hand, one may apply
finite-size scaling analysis to estimate critical properties.

Let ${\bar p_c} \equiv \lim_{t \to \infty} p_c(t)/P(t)$ denote the QS probability of
configuration $c$, where $p_c(t)$ is the probability at time $t$
and $P(t)$ is the survival probability, i.e., that probability that the absorbing state
has not been visited up to time $t$.
The QS distribution is normalized so: $\sum_c {\bar p_c} = 1$, where the sum is over
{\it nonabsorbing} configurations only (the QS probability of any absorbing
configuration is zero by definition).
Given the set of all configurations (including absorbing ones) and the values
of all transition rates ${w}_{c',c}$ (from $c$ to $c'$), we construct the QS distribution
via the iterative scheme demonstrated in \cite{analise numerica}:

\begin{equation}
\label{iterar}
{\bar p}'_c=a \bar{p}_c+(1-a) \frac{r_c}{w_c-r_a}
\end{equation}

\noindent Here $r_c=\sum_{c'} \textsc{w}_{c,c'}{\bar p}_{c'}$ is the
probability flux (in the master equation) into state $c$, $r_a$ is
the flux to the absorbing state ($1/r_a$ gives the lifetime of QS
state), and $w_c=\sum_{c'}\textsc{w}_{c',c}$ is the total rate of
transitions out of state $c$. The parameter $a$ can take any value
between $0$ and $1$ (in practice we use $a=0.1$).  Following each
iteration, the resulting distribution ${\bar p}'_c$ is normalized by
multiplying each probability by $f = 1/[\sum_c {\bar p}'_c]$.
Starting from an initial guess (for example, a distribution uniform
on the set of nonabsorbing configurations), this scheme rapidly
converges to the QS distribution.

Since the number of configurations and transitions grows very
rapidly with system size, we use a computational algorithm for their
enumeration. To begin, we enumerate all configurations of $L/2$
particles on a ring of $L$ sites (recall that each particle must
occupy a distinct site). Configurations that differ only by a
lattice translation or reflection are treated as equivalent.  Thus
the space of configurations is divided into equivalence classes
${\cal C}$.  For each class we store one representative
configuration, and the number $|{\cal C}|$ of configurations in the class,
which we call its {\it weight}. Each
configuration is determined by (1) the particle positions and (2)
the state (active or sleeping) of each particle.  If we ignore the
particle states, the particle positions define the {\it basic
configuration}; each basic configuration corresponds to a series of
configurations $c$.  One such configuration has all particles
active, while others have $1, 2,...,L/2$ inactive particles;  the
one with all particles inactive is absorbing.  Once the set of basic
classes has been enumerated, we enumerate the classes with
$n_p=0,1,...,L/2$, inactive particles, and their associated weights.

Next, we enumerate all transitions between configurations.
We visit each equivalence class ${\cal C}$
in turn, and enumerate all the manners in which ${\cal C}$ arises
in a transition (due to particle hopping, inactivation, or activation)
from an antecedent configuration in some class ${\cal C}'$.
Each transition is characterized by a rate $w_{{\cal C},{\cal C}'}$
and by an associated weight, $m_{{\cal C},{\cal C}'}$.  (The latter
is needed because in certain cases, two or more distinct transitions to the same
class ${\cal C}$ have antecedent configurations belonging to the same
class, ${\cal C}'$.)  Given the set of classes and transitions, and associated
rates and weights, we can iterate the relation given above to determine the
QS probability distribution on the set of nonabsorbing classes.  (The sums are
now over classes, with normalization taking the form
$\sum_{\cal C} |{\cal C}| \, {\bar p}_{\cal C} = 1$.)

We determine the QS distribution on rings of size $L=6, 8, 10, ... , 22$. For
$L=22$ the total number of equivalence classes is $N_{conf}=32\,842\,718$, and
the number of transitions
involving hopping and sleeping are $N_h=265\,512\,131$
and $N_s=180\,594\,624$, respectively.  Our criterion for convergence of Eq.
(\ref{iterar}) is that the sum of all absolute differences
between the probabilities ${\bar p}_{\cal C}$ and ${\bar p}_{\cal C}'$ at successive
iterations be smaller than $10^{-15}$.

\subsection{Critical properties}

Extracting estimates for critical properties from results for small
systems depends on finite-size scaling (FSS)
analysis \cite{fss1,fss2}.
The FSS hypothesis implies that the order parameter
follows $\rho(\Delta, L) \propto L^{-\beta/\nu_\perp} {\cal
R}(L^{1/\nu_\perp} \Delta)$, where
$\Delta \equiv (\lambda_c - \lambda)/\lambda_c$ and ${\cal R}$ is a scaling function.
(Note that in the SRW model the active phase corresponds to $\lambda < \lambda_c.$)
The QS order
parameter is given by $\rho=(L/2)^{-1} \sum_c {\bar p}_c N_{a,c}$,
with $N_{a,c}$ the number of active particles in configuration $c$.
To find the critical exponent $\tilde{\beta} \equiv \beta/\nu_\perp$, we seek
crossings of the quantities \cite{mendonca},

\begin{equation}\label{sl} S_L(\lambda)\equiv
\frac{\ln[\rho(\lambda,L+1)/\rho(\lambda,L-1)]}{\ln[(L+1)/(L-1)]},
\end{equation}
\vspace{.2em}

\noindent for successive pairs of system sizes.
Let  $S_{L+1}(\lambda) = S_{L-1}(\lambda) \equiv
\tilde{\beta}(L)$ for $\lambda = \lambda_{S,L}$. The crossing
values $\lambda_{S,L}$ and $\tilde{\beta}(L)$ are expected to converge to
$\lambda_c$ and $\tilde{\beta}$, respectively, as $L\rightarrow\infty$.

To estimate the dynamic exponent $z$, we determine the QS probability
flux to the absorbing state (i.e., the inverse lifetime),
which follows
$r_a \propto L^{-z} {\cal F}(\Delta L^{1/\nu})$, with ${\cal F}$ another scaling
function. The crossings of

\begin{equation}\label{ra}
R_L(\lambda)\equiv
\frac{\ln[r_a(\lambda,L-1)/r_a(\lambda,L+1)]}{\ln[(L+1)/(L-1)]}
\end{equation}
\vspace{.2em}

\noindent furnish a series of estimates, $z_L$.  As in the case of $S_L$ above, the $\lambda$ values,
$\lambda_{R,L}$,
at the crossings are expected to converge to $\lambda_c$.

Critical behavior at an absorbing state phase transition is also
characterized by order-parameter moment ratios \cite{rdjaff}.
Let $m_k$ denote the $k$-th moment of the order parameter.
The scaling property of the QS probability distribution leads to the
asymptotic size-invariance of moment ratios of the form
$m_n/(m_r^im_s^j)$ for $ir+js=n$, at the critical point.
(Although not, strictly speaking, a {\it ratio}, the product
$m_{-1} m$ of the first positive and negative moments follows the same
general scheme.)
We analyze
the ratios $m_{211} \equiv m_2/m_1^2$, $m_{3111} \equiv m_3/m_1^3$,
$m_{-1} m$,
and the reduced fourth cumulant, or kurtosis $q$.  The latter is defined so:
$q=K_4/K_2^2$, where $K_2=m_2-m_1^2 =$ var$(\rho)$ and
$K_4=m_4-4m_3m_1-3m_2^2+12m_2m_1^2-6m_1^4$. The $\lambda$ values
marking the crossings of the moment ratios (for system sizes $L$ and
$L+2$) are once again expected to converge to $\lambda_c$ as $L \to
\infty$. The values of the moment ratios and $q$ at the
critical point are universal quantities, determined by the scaling
form of the order-parameter probability distribution
\cite{binder,rdjaff}, and so are useful in identifying the universality class.

A further quantity of interest is the scaled
variance of the order parameter $\chi =L^d (\bar{\rho^2}-\bar{\rho}^2)$
which is expected to diverge as $|\lambda - \lambda_c|^{-\gamma}$.
(In equilibrium systems, $\chi$
is proportional to the susceptibility).
In a system of size $L$,
$\chi$ exhibits a maximum at a sleeping rate we denote $\lambda_{\chi,L}$.
FSS predicts that at
the critical point $\chi \propto L^{\gamma/\nu_\perp}$.

While the quantities mentioned above furnish the exponent {\it ratios}
$\beta/\nu_\perp$, $\nu_{||}/\nu_\perp \equiv z$, and $\tilde{\gamma} \equiv \gamma/\nu_\perp$, it is also
possible to estimate $\nu_\perp$ directly. FSS
implies that $m_{211} \simeq {\cal R}(\Delta L^{1/\nu_\perp})$,
where ${\cal R}$ is a scaling function. Thus
$r' \equiv |d m_{211,L}/d \lambda |_{\lambda_c} \propto L^{1/\nu_\perp}$.
The derivatives of other moment ratios, of $\ln \rho$, and of $\ln r_a$ scale
in an analogous manner.

Given a series of estimates for the critical point (or for
a critical exponent, or a moment ratio), associated with a sequence of sizes $L$, we extrapolate
to infinite size using polynomial fits and the Bulirsch-Stoer (BST)
procedure \cite{bst}.
In the contact process \cite{exact sol},
quantities such as $\lambda_{S,L}$ and $\tilde{\beta}(L)$
vary quite systematically with system size, leading to precise
estimates for critical values via BST extrapolation.

\begin{figure}[!h]
\includegraphics[width=12cm]{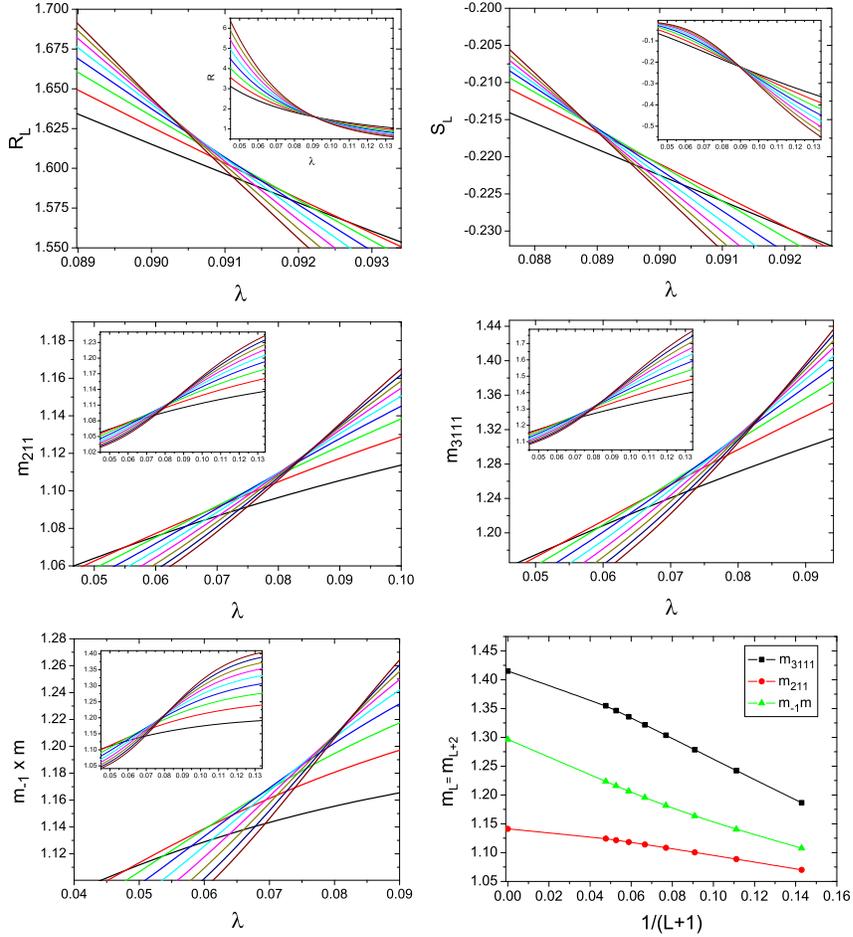}
\caption [no cap]{QS analysis: $R$, $S$, $m_{211}$, $m_{3111}$,
$m_{-1}m$ in the neighborhood the crossings.
The insets show these quantities over a larger range of $\lambda$ values.
The lower-right panel shows the
crossing values for $m_{211}$,$m_{-1}m_1$ and $m_{3111}$ (lower to
upper) along with the extrapolated ($L \to \infty$) values.} \label{h2t}
\end{figure}

\begin{figure}[!h]
\includegraphics[width=10cm]{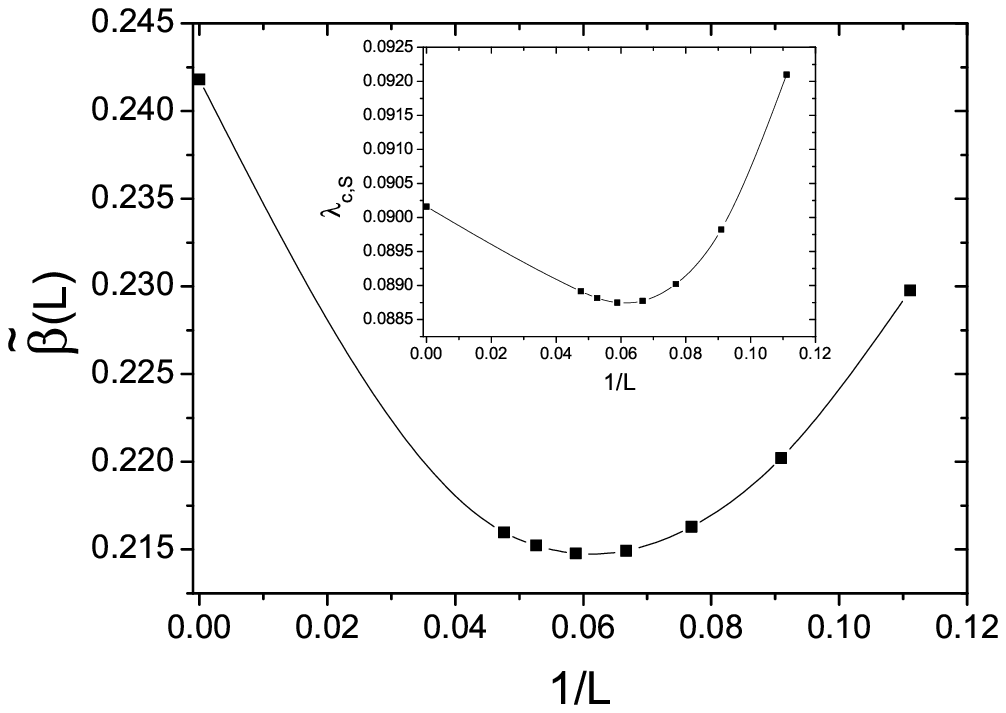}
\caption [Expoente beta] {Crossing values $\tilde{\beta}(L)$ versus
$1/L$.
The inset is a similar plot of the crossing points $\lambda_{S,L}$.
The leftmost points are the extrapolated values using BST; curves
are splines to the data and the BST extrapolations, intended as a
guide to the eye. Error bars are smaller than the symbols.}
\label{g1}
\end{figure}

\begin{figure}[!h]
\includegraphics[width=10cm]{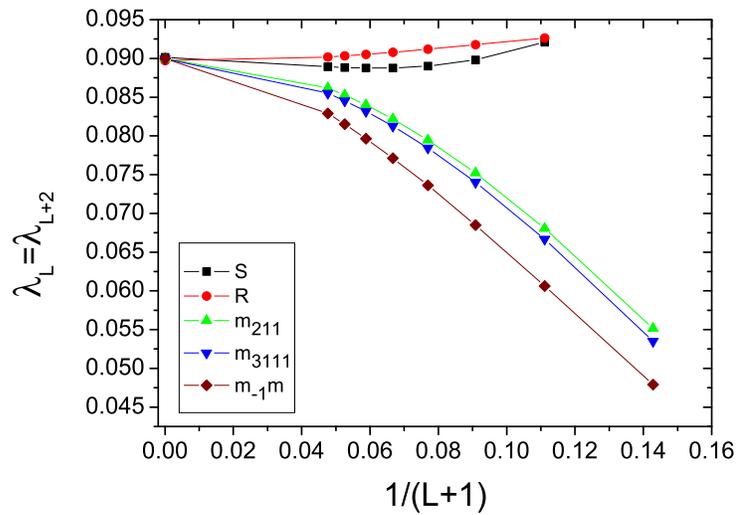}
\caption [sem cap] {Crossing values $\lambda_{.,L}$ for
$R_L$, $S_L$, $m_{211}$, $m_{3111}$ and $m_{-1}m_1$ (upper to lower).}
\label{extraprate}
\end{figure}

Our first task is to determine the critical sleeping rate $\lambda_c$;
to this end we analyze the crossings of $S$, $R$, and the moment ratios.
In a preliminary analysis the crossings
are determined graphically; the general tendencies are shown in Fig. \ref{h2t}.
Subsequently, we refine these estimates
by calculating these quantities at intervals of $\delta
\lambda=10^{-5}$, at $20$ points around each estimated crossing, and
determine the crossing values
to a precision of $10^{-12}$ or better
using Neville's algorithm \cite{algnev}. Figure \ref{g1}
shows the results for crossings of $S_L$.  This quantity is unusual in that
the crossing values are nonmonotonic; for the other quantities studied, the
$L$-dependence is monotonic over the accessible range of system sizes
(see Fig. \ref{extraprate}).

In BST extrapolation \cite{bst} the limiting value $T_\infty$ of a sequence $T_j$
($j = 1, 2, 3,...$), is estimated
on the basis of the first $N$ terms via
the recurrence relations

\begin{equation}\label{bsteq}
T_m^{(n)}=T_{m-1}^{(n+1)}+(T_{m-1}^{(n+1)}-T_{m-1}^{(n)})\left[\left(\frac{h_n}{h_{n+m}}\right)^\omega
\left(1-\frac{T_{m-1}^{(n+1)}-T_{m-1}^{(n)}}{T_{m-1}^{(n+1)}-T_{m-2}^{(n+1)}}\right)-1\right]^{-1}
\end{equation}
\vspace{.2em}

\noindent where, for $j = 1,...,N$, $T_{-1}^{(j)} \equiv 0$,
$T_0^{(j)} \equiv T_j$, and $h_j$ is a sequence converging to zero as $j
\rightarrow \infty$. (Here $h_j = 1/\bar{L}_j$, where $\bar{L}_j$ is the system
size, or, for crossings,
the mean value of the two system sizes involved.)

\begin{figure}[!h]
\includegraphics[width=10cm]{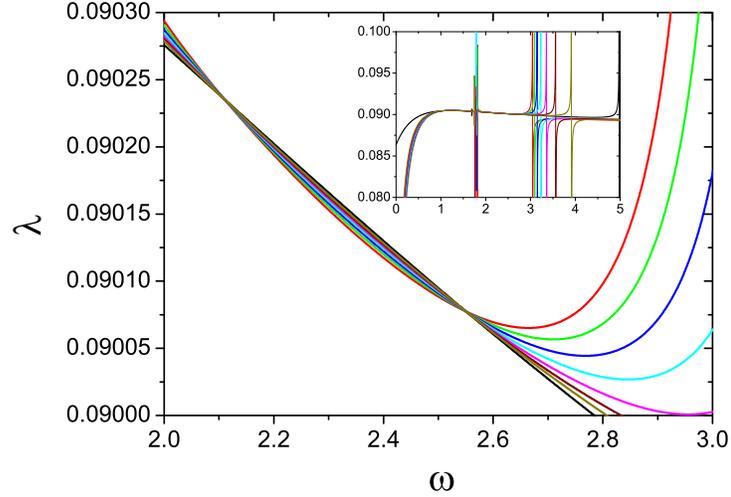}
\caption [erro em bst] {BST estimates (see text) for $\lambda_{c,S}$ versus extrapolation
parameter $\omega$.  Main graph: detail of interval [2,3]; inset: the full
interval of study.} \label{h2}
\end{figure}

The BST procedure includes a free parameter, $\omega$, which can be adjusted
to improve convergence.  We use a convergence criterion similar to
one employed in analyses of series expansions via Pad\'e
approximants \cite{privman,rdiwanjsp}, in which, varying some parameter,
one seeks concordance amongst the estimates furnished by various approximants.
In the present case, given $N$ values $T_j$, each associated
with inverse system size $h_j$, we calculate $N+1$ estimates, one using the
full set, and $N$ others obtained by removing one point, $(h_k,T_k)$, from
the set.  We search for values of $\omega$ that minimize the differences
between the various estimates.
Sweeping the interval $[0,5]$, we find that each quantity studied
($\tilde{\beta}$, $z$, moment ratios, and the associated estimates for
$\lambda_c$) exhibits one or more crossings at which all $N+1$
estimates are equal to within numerical precision.  Figure \ref{h2}, for
$\lambda_{c,S}$ (the estimate for $\lambda_c$ derived from the crossing
values $\lambda_{S,L}$), illustrates the typical behavior.
In this case there are four crossings, which
fall at $\omega$ = 1.051380, 1.658703, 2.109962, and 2.550852;
the associated values of $\lambda_{c,S}$ are 0.0904577, 0.0903878,
0.0902354, and 0.0900773.

To choose among the values when there are multiple crossings, we note
that $\omega$ in the BST procedure is effectively a correction
to scaling exponent.  An independent estimate for this exponent can
be obtained via a least-squares fit to the data using a double power-law
form, for example,

\begin{equation}
\lambda_{S,L} = \lambda_{c,S} + \frac{A}{L^{y_1}} + \frac{B}{L^{y_2}}
\label{pwrfit}
\end{equation}

\noindent with $y_2 > y_1$.  (The best-fit parameters $A$, $B$, $y_1$ and
$y_2$ are determined by minimizing the variance of the
differences $\delta_L \equiv \lambda_{S,L} - A L^{-y_1}  - B L^{-y_2}$.)
This yields $y_1 = 2.02$, leading us to take the average of the
two values associated with the BST crossings nearest $y_1$, resulting
in $\lambda_{c,S} = 0.09016$.  A similar procedure is used to obtain
the other estimates listed in Tables \ref{lambdac} and \ref{cruzbst}.
(We note that the apparent correction to scaling exponent $y_1$
falls in the range 2.02-2.23 for the crossing values of $\lambda$,
and in the range 1.1-1.6 for the associated quantities $\tilde{\beta}$,
z, and the moment ratios.)

Tables \ref{lambdac} and \ref{cruzbst} include polynomial extrapolations as alternative
estimates for the quantities of interest.  (In this case we fit the data
to a polynomial in $1/L$, using the highest possible degree.  Polynomials of
degree one or two smaller than maximum yield very similar results.)
We adopt the mean of the BST and polynomial extrapolations as our best estimate,
and adopt the difference between the two results
as a rough estimate of the associated uncertainty.  The situation is
particularly favorable for determining $\lambda_c$ since we have five independent
estimates.  The average of the BST results is $\lambda_c = 0.08996(7)$ while that
from the polynomial fits is 0.09007(10), leading to our best estimate
of $\lambda_c = 0.09002(10)$.  (Figures in parentheses denote uncertainties,
given as one standard deviation.)  The estimates for $\lambda_c$ and the moment ratios
are consistent with simulation results, whereas those for $\beta/\nu_\perp$ and $z$ are not.
(A detailed comparison is given in the following section.)
Figure \ref{extraprate}
shows the finite-size data and BST extrapolations for the various crossing
values, $\lambda_{.,L}$.

\begin{figure}[!h]
\includegraphics[width=10cm]{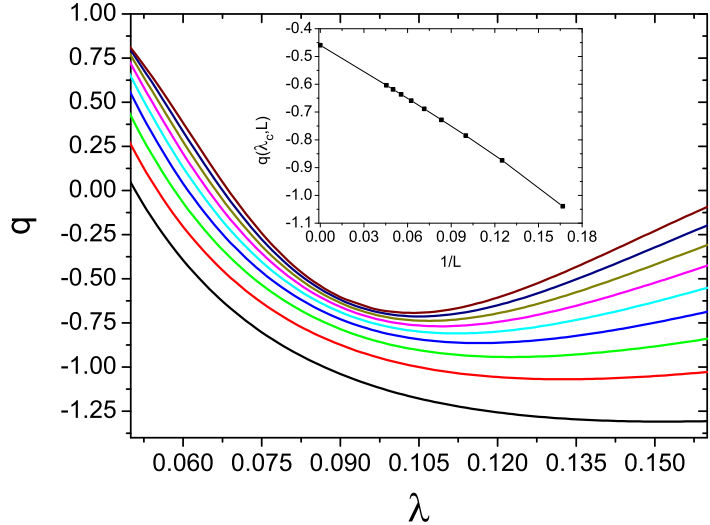}
\caption [sem cap] { Kurtosis $q$ vs $\lambda$ for
sizes $6$, $8$, ..., $22$ (lower to upper). Inset:
values for $q(L)$ at the critical point and our estimate for $q_{\lambda_c, \infty}$.}
\label{kurt}
\end{figure}

\begin{figure}[!h]
\includegraphics[width=10cm]{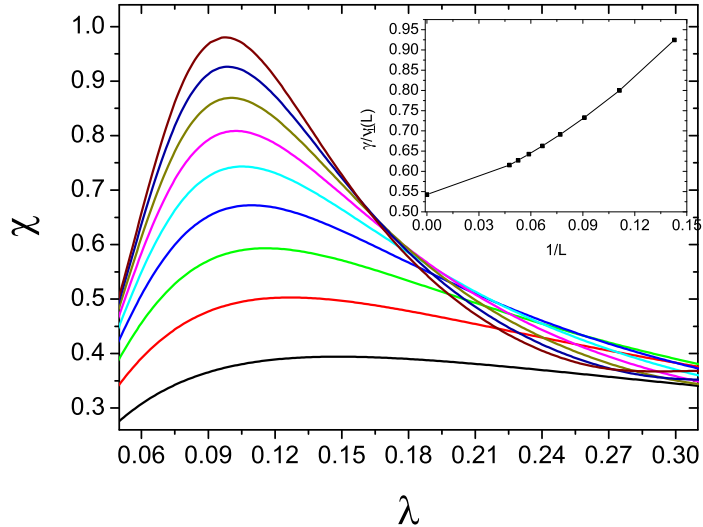}
\caption [sem cap] {Scaled order parameter variance $\chi$ versus
$\lambda$ for sizes $6$, $8$, ..., $22$ (lower to upper). Inset:
Estimates for $\gamma/\nu_\perp$ obtained via local-slopes analysis,
and the extrapolated (infinite-size) value.} \label{suscept}
\end{figure}

The values for the kurtosis at $\lambda_c$
approach a limit of $q_c = -0.454(2)$ (see Fig. \ref{kurt}).
For a given system size, $q(\lambda,L)$ exhibits a minimum in the vicinity of
$\lambda_c$, as also observed in the contact process \cite{exact sol}.
Extrapolating $q(\lambda,L)$ to $L \to \infty$ for a series of values near
$\lambda_c$, we obtain a function that exhibits a minimum near $\lambda = 0.0947$
(the minimum value is -0.521).  This implies that the minimum falls near, but not
at the critical point, a conclusion supported by the simulation data reported
in the following section.
Turning to the scaled order-parameter variance $\chi$, we observe
pronounced maxima even in small systems, as illustrated
in Fig. \ref{suscept}.
Estimates for $\tilde{\gamma}$ obtained from a local-slopes analysis of
$\chi$ at the critical point $\lambda_c$ are listed in Table \ref{cruzbst}.
(The local slope is defined so: $\tilde{\gamma}(L) \equiv \ln[\chi(L+1)/\chi(L-1)]/\ln[(L+1)/(L-1)]$.)

\begin{table}[h]
\begin{minipage}[b]{1.\linewidth}
\centering \caption[tabel2]{Estimates for the critical sleeping rate
$\lambda_c$
obtained via analysis of the QS probability distribution
using BST extrapolation and polynomial fits.} \label{lambdac}
\begin{center}
\begin{tabular}{|c|l|l|}
\hline
   Quantity        &   BST   & Polynomial \\ \hline \hline
$\lambda_{c,S}$    & 0.09016 &  0.09039   \\
$\lambda_{c,R}$    & 0.08973 &  0.08993   \\
$\lambda_{c,211}$  & 0.08999 &  0.09008   \\
$\lambda_{c,3111}$ & 0.08995 &  0.09016   \\
$\lambda_{c,-11}$  & 0.08998 &  0.08979   \\
\hline
\end{tabular}
\end{center}
\end{minipage}
\end{table}
\vspace{2em}

\begin{table}[h]
\begin{minipage}[b]{1.\linewidth}
\centering \caption[tabel2]{Estimates for critical properties
obtained via analysis of the QS probability distribution
using BST extrapolation and polynomial fits.} \label{cruzbst}
\begin{center}
\begin{tabular}{|c|l|l|l|}
\hline
Quantity           &   BST  & Polynomial & Best Est.   \\ \hline \hline
$\beta/\nu_\perp$  & 0.2418 &  0.2405    &  0.241(1)   \\
$z$                & 1.668  &  1.660     &  1.664(4)   \\
$\gamma/\nu_\perp$ & 0.5428(1) & 0.53942(1) & 0.541(2)   \\
$m_{211}$          & 1.1412 &  1.1422    &  1.1417(5)  \\
$m_{3111}$         & 1.4151 &  1.4249    &  1.420(5)   \\
$m_{-1} m $        & 1.2965 &  1.3203    &  1.308(12)  \\
$q$                &-0.460(5) & -0.454(4) & -0.457(6)   \\
\hline
\end{tabular}
\end{center}
\end{minipage}
\end{table}
\vspace{2em}

To estimate $\nu_\perp$, we determine $r' \equiv |d m_{211,L}/d
\lambda |_{\lambda_c}$ by constructing linear fits to the data on
the interval $0.08992 \leq \lambda \leq 0.09012$, using an increment
of $\Delta \lambda =10^{-5}$. Since
the graph of $\ln r'$ versus $\ln L$ shows significant curvature, we
analyze the local slopes, $\nu_\perp (L) =
\ln[(L-1)/(L+1)]/\ln(r'_{L-1}/r'_{L+1})$. BST extrapolation of the
latter yields $\nu_\perp=1.293(5)$. We obtain independent
estimates for $\nu_\perp$ using the order parameter and flux of
probability to the absorbing state, $r_a$, as described above,
yielding 1.285(10) and 1.2644(2), respectively. On the basis of
these results, we estimate $\nu_\perp=1.28(1)$.

\section{Monte Carlo Simulations}

\subsection{Simulation methods}

We perform extensive simulations of the SRW model using both
conventional and quasistationary (QS) methods.
Quasistationary simulations \cite{howtosimulate,quasidistr} have
proven to be an efficient method for studying
absorbing state phase transitions, allowing one to obtain
results of a given precision with an order of magnitude less CPU
time than in conventional simulations.  The method samples the QS
distribution defined in the preceding section using a list of
configurations saved during the evolution; when a visit to the
absorbing state is imminent, the system is instead placed in a
configuration chosen at random from the list.  A detailed
explanation of the method is given in \cite{howtosimulate}.

We perform QS simulations using system
sizes $L=100$, $200$, $400$,..., $32000$. Each realization of the
process runs for $T=10^9$ time units, with the first $T_R =
10^8$ time units discarded to ensure all transients have been
eliminated.  (Our time unit is defined below.)
We use $1000$ saved configurations; the replacement
probability (i.e., for replacing one of the configurations on the
list with the current one) is $p_{rep}=10^{-5}.$ Our choice of
$p_{rep}$ is guided by the condition $T > \tau_M > \tau$, where
$\tau_M=M/p_{rep}$ is the mean time that a configuration remains on
the list and $\tau$ is the mean lifetime in the QS state. The latter
is estimated as $\tau=(T-T_R)/N_{abs}$, where $N_{abs}$ is the
number of (attempted) visits to the absorbing state for $t>T_R$.
During the initial relaxation period ($t < T_R$) we use
$p_{rep}=10^{-2}$ to eliminate the memory of the initial
configuration. For each value of $\lambda$ studied, we calculate the
mean and statistical uncertainties (given as one standard deviation)
over $N_R=20$ independent realizations; for $L=16000$ we use
$N_R=40$.

At each step of the simulation we select the
particle involved from a list of active particles. The time
increment associated with each step is $\Delta t=1/N_{a}$, where
$N_a$ is the number of active particles just prior to the event. For
$t > T_R$, we accumulate a histogram $h(N_a)$ of the time
during which there are exactly $N_a$ active particles. The
normalized histogram is our best estimate for the probability
distribution $P(N_a)$, from which we may determine any desired
moment of the order parameter $\rho= N_a/N$. The QS lifetime $\tau$
may also be obtained from $P(N_a)$ via the relation $\tau_h =
1/[\lambda P(1)]$, where the subscript $h$ serves only to
distinguish this from the value $\tau$ found using the mean time between visits to
the absorbing state.

\subsection{Scaling at the critical point}

\begin{figure}[!h]
\includegraphics[width=10cm]{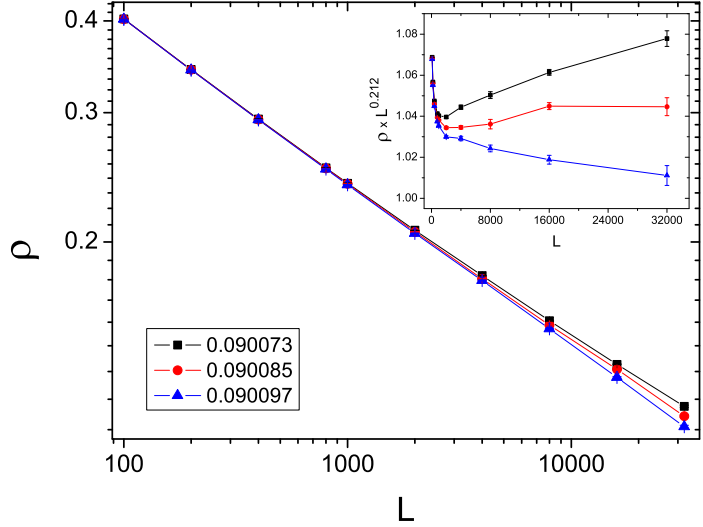}
\caption [cruzm] {Order parameter $\rho$ versus system size $L$.
Inset: $L^{0.212} \rho$ versus $L$. Lines are a guide to the
eye. } \label{mc1}
\end{figure}

\begin{figure}[!h]
\includegraphics[width=10cm]{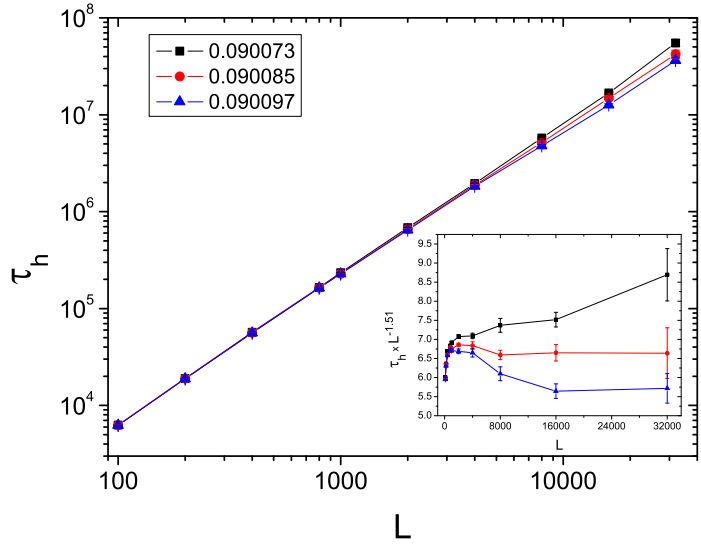}
\caption [cruzm] {Lifetime $\tau_h$ versus system size $L$. Inset:
$L^{-1.51} \tau_h$ versus $L$. Lines are a guide to the eye.}
\label{tauhf}
\end{figure}

To determine $\lambda_c$, we first locate the crossings of the
moment ratios $m_{211}$ (defined in Sec. III), for pairs
of consecutive system sizes, and obtain a preliminary estimate by
extrapolating the crossing values to $L \to \infty$.  We then study
larger systems using $\lambda$ values close to our preliminary
estimate.  Using these results, we determine the critical value via
the familiar finite-size scaling criteria $\rho \sim
L^{-\beta/\nu_\perp}$ and $\tau \sim L^z$, and the condition that
$m_{211}$ approach a finite limiting value as $L$ increases.  In
logarithmic plots, $\rho$ and $\tau$ exhibit upward (downward)
curvature for $\lambda < \lambda_c$ ($> \lambda_c$) as illustrated
in Figs. \ref{mc1} and \ref{tauhf}; off-critical values are also
readily identified in plots of $m_{211}$ versus $1/L$. Using these
criteria we obtain $\lambda_c = 0.090085(12)$.
Of note are the strong finite-size corrections (evident in the
insets of Figs. \ref{mc1} and \ref{tauhf}), for $L < 1000$. Indeed,
our estimates for critical exponents and moment ratios are obtained
using only the data for $L \geq 1000$.

We turn now to FSS estimates of critical exponents.
Using the data for $L
\geq 1000$ we obtain $\beta/\nu_\perp = 0.212(6)$ from analysis of
$\rho$, and 0.217(10) from analysis of $m_{-1}$, which as noted
in Sec. III, is expected to follow
$m_{-1}(L,\lambda_c) \propto L^{\beta/\nu_\perp}$.
Analysis of the QS
lifetime using $\tau$ and $\tau_h$ yields $z=1.50(4)$ and $z=1.51(4)$
respectively, while the data for the moment ratio yield
$m_{211,c}=1.141(8)$. Restricting the analysis to the data for $L
\geq 2000$, or for $L \geq 4000$, yields estimates for
$\beta/\nu_\perp$, $m_{211,c}$, and $z$ consistent with
the values cited above, but with somewhat larger
uncertainties. Analysis of $\chi$ yields $\gamma/\nu_\perp = 0.58(1)$. In all cases the chief
contribution to the uncertainty is due to the uncertainty in
$\lambda_c$ itself.

\begin{figure}[!h]
\includegraphics[width=10cm]{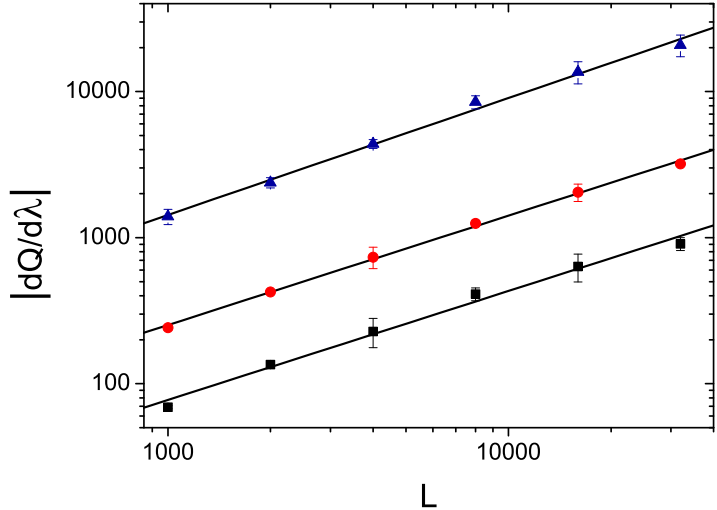}
\caption [nuperp] {Derivatives $r' \equiv |d Q/d \lambda
|_{\lambda_c}$ for $m_{211}$ (squares), $\ln \rho$ (circles) and
$\ln \tau$ (triangle). Lines are linear fits to the data, with
slopes of $0.746(37)$, $0.749(16)$, and $0.80(3)$ (lower to upper).}
\label{nup}
\end{figure}

To estimate the exponent $\nu_\perp$ directly, we apply the method
used in Sec. \ref{secQS}, calculating the derivatives $r'$ of
quantities such as $\ln \rho$ and $\ln \tau $ with respect to
$\lambda$ near the critical point.  Using simulation data for
$\lambda=0.090073$, $0.090085$, and $0.090097$, we construct a
linear fit to estimate $r'$ at $\lambda_c$. In Fig \ref{nup} we plot
the values for $r'$ obtained via analysis of $|d \ln \rho / d
\lambda|$, $d m_{211} / d \lambda$ and $|d \ln \tau / d \lambda|$.
The estimates obtained using the data for $L \geq 1000$ are listed
in Table \ref{nuperpmc}; based on these results we estimate
$\nu_\perp=1.31(4)$.  Since the values obtained using different
quantities are quite different, this exponent is not determined to
good precision.

\begin{table}[h]
\begin{minipage}[b]{1.\linewidth}
\begin{footnotesize}
\centering \caption[sem cap]{\footnotesize Estimates for $\nu_\perp$
obtained from the derivatives of $m_{211}$, $\ln \rho$ and $\ln \tau
$.} \label{nuperpmc}
\end{footnotesize}
\begin{center}
\begin{tabular}{p{4cm} p{2.7cm} p{2.7cm} p{2.7cm}}
\hline \hline
   $r' \equiv |d Q/d \lambda
|_{\lambda_c}$ &  $Q \equiv m_{211}$  &  $Q \equiv \ln(\rho)$ &  $Q \equiv \ln(\tau)$  \\
\hline $1000 \leq L \leq 32000$ &  $1.34(7)$ & $1.34(4)$ & $1.25(5)$
\\ \hline
\end{tabular}
\end{center}
\end{minipage}
\end{table}

\begin{figure}[!h]
\includegraphics[width=10cm]{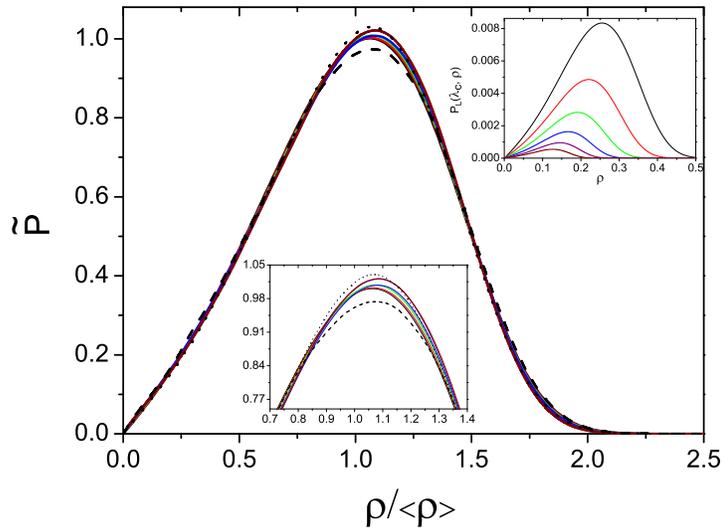}
\caption [cruzm] {Main figure: scaling function ${\tilde P} \equiv
(\langle \rho \rangle L/2) P_L (\rho/\langle \rho \rangle)$ versus
$\rho/\langle \rho \rangle$ for system sizes $L=1000$, $2000$,...,$L=32\,000$
(lower to upper). The dotted and dashed curves
show the corresponding result for the one-dimensional restricted stochastic sandpile,
for $L=20\,000$ and $50\,000$, respectively.
Lower inset: detail of the region in which
$\tilde{P}$ takes its maximum. Upper inset: unscaled data (system
sizes increasing from upper to lower).}
\label{histreesc}
\end{figure}

Next we examine the QS probability distribution $P_L
(\rho,\lambda)$ for the fraction of active particles, $\rho =
N_a/N$. At the critical
point, this distribution is expected to take the scaling form
\cite{densprob},
\begin{equation}\label{reshist}
P_L(\rho, \lambda_c)= \frac{2}{\langle \rho \rangle L}
\tilde{P}(\rho/\langle \rho \rangle),
\end{equation}
where $\tilde{P}$ is
a normalized scaling function. (The prefactor arises from
normalization of $P_L$, with $N=L/2$.) Figure \ref{histreesc} is a
scaling plot of the QS probability distribution at the critical
point, showing evidence of a data collapse, albeit with significant
finite-size corrections (the maximal scatter of the values is about 2\%);
the collapse is quite good for the two
largest system sizes.  Also shown are scaled probability distributions for
the one-dimensional conserved restricted stochastic sandpile \cite{sandpile2006}, showing
good overall agreement.

\begin{figure}[!h]
\includegraphics[width=10cm]{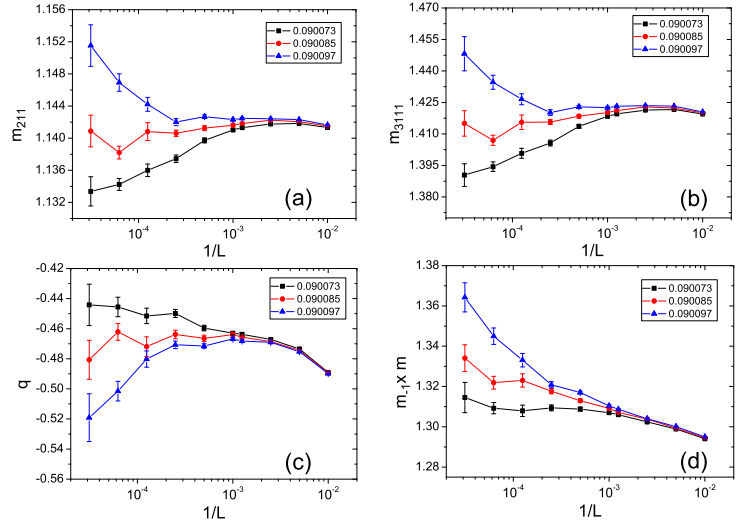}
\caption [cruzm]{Moment ratios (a) $m_{211}$; (b) $m_{3111}$; (c)
reduced fourth cumulant $q$; and (d) $m_{-1}m $ versus system size
$L$.} \label{tdmom}
\end{figure}

Figure~\ref{tdmom} shows the behavior of the
order-parameter moment ratios $m_{211}$, $ m_{3111}$, the reduced
fourth cumulant $q$, and $m_{-1}m$, defined in Sec. III.  Our estimates for
the critical values are given in Table \ref{compQSsim}).

\subsection{Off-critical scaling behavior}

It is of interest to study the scaling of the order parameter, of the
scaled variance and of the lifetime, away from the critical point.
Although off-critical scaling properties have been amply
verified for models in the directed percolation universality class,
such as the contact process \cite{densprob,livrodickman}, finite-size scaling
and associated data collapse of the order parameter is more
problematic in one-dimensional stochastic sandpile models
\cite{sandpile2002,sandpile2006}.

FSS analysis implies that the order parameter
take the form $\rho(\Delta, L) \sim L^{-\beta/\nu_\perp} f(\Delta
L^{1/\nu_\perp})$, where the scaling function is $f(x)\propto
x^\beta$ for $x \gg 1$.
In the inactive phase, $\lambda >\lambda_c$, the number of active particles in the QS regime is
${\cal O}(1)$, so that $\rho \propto L^{-1}$, leading to $f(x) \sim
|x|^{\beta-\nu_\perp}$.
Alternatively, writing
$\rho(\Delta, L) \sim \Delta^\beta h(\Delta L^{1/\nu_\perp})$,
the scaling function must satisfy $h(x)\propto x^{-\beta}$ for $x \rightarrow 0$,
while in the inactive phase $h(x) \propto |x|^{-\nu_\perp}$.
FSS analysis predicts that the scaled variance take the form
$\chi (L,\Delta) \sim L^{\gamma/\nu_\perp} g(\Delta L^{1/\nu_\perp})$,
with $g(x)\propto x^{-\gamma}$ for $x \gg 1$ and
$g(x) \propto |x|^{-(\nu_\perp+\gamma)}$ in the inactive
phase.

\begin{figure}[!h]
\includegraphics[width=14cm]{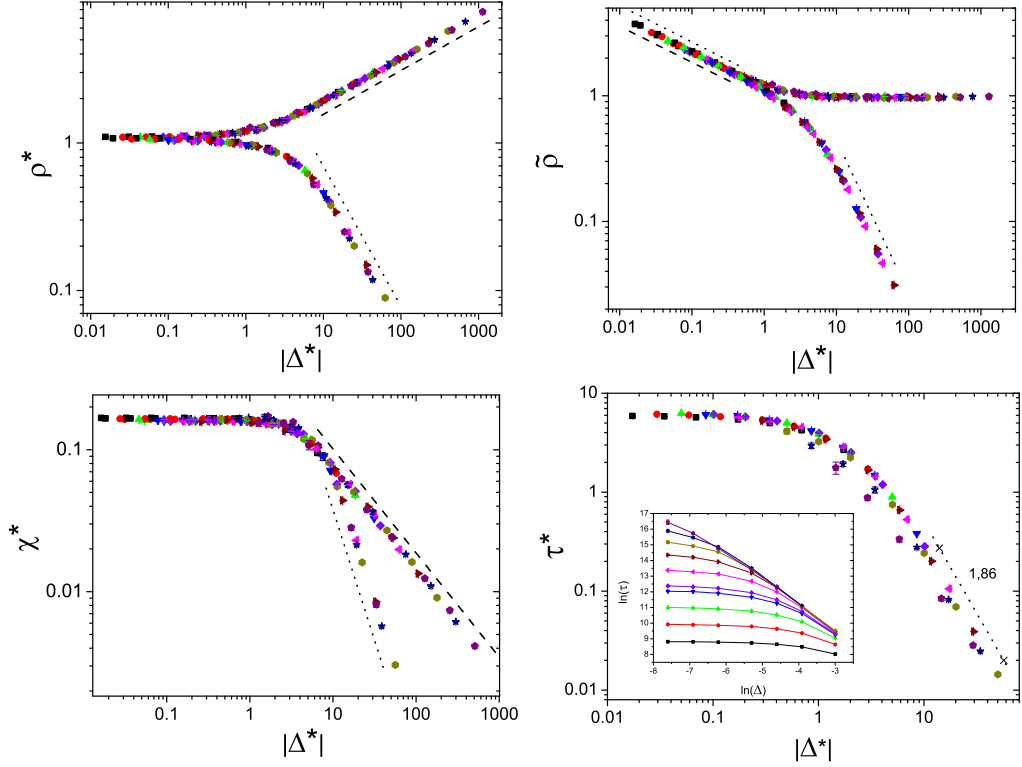}
\caption [cruzm]{Scaled order parameter $\rho^* \equiv \rho
L^{\beta/\nu_\perp}$ (upper left) and $\tilde{\rho} \equiv \rho
\Delta^{-\beta}$ (upper right), scaled variance $\chi^* \equiv \chi
L^{-\gamma/\nu_\perp}$ (lower left) and the scaled lifetime
$\tau^* \equiv L^{-z}\tau$ (lower right) versus scaled distance from
critical point, $\Delta^* \equiv \Delta L^{1/\nu_\perp}$, in the active and
inactive phases (upper and lower set of points, respectively). The best-fit
exponents associated with the data collapses are given in Table \ref{bestexp}.
The slopes of the dashed and dotted lines represent the power
laws exhibited by the scaling functions in the active and inactive phase,
respectively.} \label{tdcol}
\end{figure}

Figure \ref{tdcol} shows a good data collapse of
the order parameter in the forms
$\rho^* \equiv L^{\beta/\nu_\perp} \rho$ and
$\tilde{\rho} \equiv \rho \Delta^{-\beta}$, and of the scaled order-parameter variance
$\chi^* \equiv \chi L^{-\gamma/\nu_\perp}$, as functions of $\Delta^* \equiv \Delta
L^{1/\nu_\perp}$, using data for system sizes from $L=100$ to $32000$.
The exponents associated with the best
collapses are listed in Table \ref{bestexp}. Based on these results,
we estimate $\beta/\nu_\perp=0.216(4)$, $\beta=0.29(1)$ and
$\gamma/\nu_\perp=0.57(1)$. We note that the values found for
$\nu_\perp$ in the active and inactive regimes differ slightly,
leading to best estimate $\nu_\perp=1.30(4)$.

In the active phase, $\lambda < \lambda_c$, a data collapse of
$\rho^*$ versus $\Delta^*$ is obtained over about four orders of
magnitude in $\Delta^*$.  Interestingly, such a data collapse is
only observed over a much smaller interval - about one order of
magnitude - in stochastic sandpiles for a comparable range of
lattice sizes \cite{sandpile2006,sandpile2009}. A linear fit to the
data (using all sizes) for $\Delta^*> 10$ yields $\beta=0.293(2)$.
Using $\tilde{\rho}$ versus $\Delta^*$ we found $\beta=0.288(3)$,
including the data for $\Delta^*<0.2$. The data for
$\chi^*$ (for $\Delta^*>10$) yield
$\gamma=0.733(7)$. The power laws associated with these exponents
are represented in Figure \ref{tdcol} by dashed lines. Using the
hyperscaling relation $\gamma=d\nu_\perp-2\beta$, the latter results
and the exponents used in the collapses in the active phase, one
finds (a) $\beta=(1.32(3)-0.733(7))/2=0.29(2)$; (b)
$\beta/\nu_\perp=(1-\gamma/\nu_\perp)/2=(1-0.568)/2=0.216(5)$; and
(c) $\nu_\perp=\gamma+2\beta=0.733(7)+0.580(8)=1.31(2)$.  These
predictions are consistent with the values used in the collapses
and with those obtained at the critical point.

\begin{table}[h]
\begin{minipage}[b]{1.\linewidth}
\centering \caption[tabel2]{Critical exponent estimates from off-critical simulations.  For each
scaling relation (first column), we list the associated exponent obtained via a fit to the
data (second column).  The third and fourth columns give the exponents obtained via data collapse.
$x$ denotes the argument of the relevant scaling function.}
\label{bestexp}
\begin{center}
\begin{tabular}{|c|l|l|l|}
\hline
\multicolumn{4}{|c|}{Active phase} \\ \hline
$\rho^*\propto x^\beta$          & $\beta$ = 0.293(2) & $\beta/\nu_\perp$ = 0.218(5) & $\nu_\perp$ = 1.34(3) \\
\hline
$\tilde{\rho}\propto x^{-\beta}$ & $\beta$ = 0.288(3) & $\beta$ = 0.293(3)           & $\nu_\perp$ = 1.32(3) \\
\hline
$\chi^*\propto x^{-\gamma}$     & $\gamma$ = 0.733(7) & $\gamma/\nu_\perp$ = 0.568(8)& $\nu_\perp$ = 1.32(3) \\
\hline\hline
\multicolumn{4}{|c|}{Inactive phase} \\ \hline
$\rho^* \sim x^{\beta-\nu_\perp}$     & $\beta-\nu_\perp = -0.95(4)$ & $\beta/\nu_\perp$ = 0.214(2) &
$\nu_\perp$ = 1.26(2) \\
\hline
$\tilde{\rho}\sim x^{-\beta}$         & $\beta = 0.29(1)$ & $\nu_\perp = 1.22(5)$ & $\nu_\perp$ = 1.26(3) \\
\hline
$\chi^*\sim x^{-(\nu_\perp+\gamma)}$  & $\nu_\perp+\gamma = 1.85(4)$ & $\gamma/\nu_\perp$ = 0.57(1) &
$\nu_\perp$ = 1.28(3) \\
\hline
$\tau^* \sim x^{-\nu_\parallel}$ & $\nu_\parallel = 1.86(6)$ & $\nu_\parallel/\nu_\perp$ = 1.53(3) &
$\nu_\perp$ = 1.30(2) \\
\hline
\end{tabular}
\end{center}
\end{minipage}
\end{table}
\vspace{2em}

Similarly, the scaled quantities $\rho^*,\, \tilde{\rho}$ and
$\chi^*$ exhibit a good collapse in the
inactive phase, for all system sizes studied, as shown in Fig. \ref{tdcol}.
Using $\rho^*$, a linear fit to the data for $|\Delta^*|
>20$ yields $\nu_\perp - \beta = - 0.95(4)$. Using the result from
the best collapse we have $\beta=0.95(4)-1.26(2)=0.29(4)$. Analyzing the
data collapse for $\tilde{\rho}$ we find $\beta=0.29(1)$ in the small-$|\Delta^*|$ regime.
In the opposite limit we
obtain $\nu_\perp=1.22(5)$ as illustrated in the dotted lines in
Fig. \ref{tdcol}. Finally, the data collapse of $\chi^*$ versus
$\Delta^*$ leads to $\nu_\perp+\gamma=1.85(4)$.
Combining these results, the hyperscaling relation used above, and
the values from the best collapses shown in Table \ref{bestexp} for
$\chi^*$, one can predict: (a) $\gamma/\nu_\perp=0.57(1) \,
\Rightarrow \, \gamma= 0.57(1) \times 1.28(3)= 0.73(3)$, (b)
$\beta/\nu_\perp=(1-\gamma/\nu_\perp)/2=(1-0.57(1))/2=0.215(5)$ and
(c) $\gamma=1.85(4)-1.28(3)=0.57(5)$.  The first two relations are consistent
with the exponent values obtained previously, while the third conflicts with the value
for $\gamma$ found for $\lambda \leq \lambda_c$.
This may reflect a violation of scaling, but the possibility that our study does not
probe sufficiently deep into the inactive regime, for which one expects $\chi \sim 1/L$,
cannot be discarded.

In the inactive phase, the lifetime $\tau$ is expected to follow
$\tau(\Delta,L)=|\Delta|^{-\nu_\parallel} G(\Delta
L^{1/\nu_\perp})$, with the scaling function $G(x)\propto
|x|^{-\nu_\parallel}$, implying
a data collapse if we plot $\tau^*=L^{-z}\tau$ versus $\Delta^*$. As
shown in Fig. \ref{tdcol}, the collapse here is much poorer than in the other
cases. A linear fit to the data for $L\leq
8000$ and $\Delta^*>10$ furnishes $\nu_\parallel=1.86(6)$. This
result is in concordance with the scaling relation $z =
\nu_\parallel/\nu_\perp$, if we employ the values found at the
critical point ($z=1.50(4)$ and $\nu_\perp=1.34(6)$) and the best-collapse values
($z=1.53(3)$ and $\nu_\perp=1.30(2)$); the
latter yield the value $\nu_\parallel=1.53(3) \times
1.30(2)=1.99(8)$, consistent (to within uncertainty) with the value
found via collapse.

\subsection{Approach to the quasistationary regime}

Yet another aspect of scaling at an absorbing-state
phase transition involves the {\it approach} to the QS regime,
starting from a maximally active initial condition.
The quantities of interest are the time-dependent activity density
$\rho (t)$ and the moment ratio $m_{211}(t)$.  At short times
(i.e., before the QS regime is attained), at the critical point, the
expected scaling behavior for the order parameter is $\rho \sim
t^{-\delta}f(t/\tau)$ where the scaling function $f(x)\sim
x^\delta$, for $x \gg 1$, with $\delta = \beta/\nu_\parallel$.
One expects $m_{211}-1\sim t^{1/z}$ for $t \ll \tau$.

To probe this regime we perform conventional simulations for the same system sizes as used
in QS simulations; averages are calculated over $N_R=1000$
independent realizations; each runs until the system reaches the
absorbing state or attains a maximum time, $t_{max}=10^8$. We use
the critical point value $\lambda_c=0.090085$ found in QS
simulations.  We determine $\rho (t)$ and $m_{211}(t)$ as averages
over surviving realizations on time intervals that for large $t$
represent uniform increments of $\ln(t)$, a procedure known as
logarithmic binning.

\begin{figure}[!h]
\includegraphics[width=10cm]{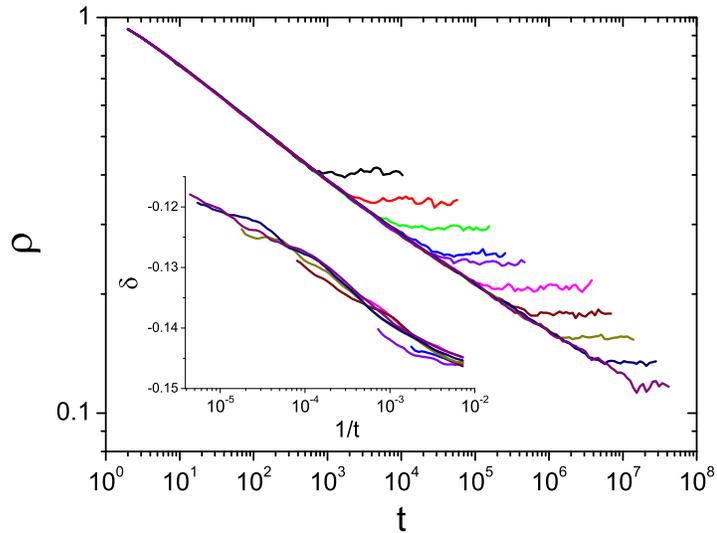}
\caption [cruzm]{Active particle density $\rho$ versus $t$. System
sizes $L=100$, $200$, $400$,...,$L=32000$ (from upper to lower). The
inset shows the local slope of $\rho (t)$ versus $1/t$.} \label{rhoaxt}
\end{figure}

The evolution of $\rho$ as a function of $t$ is shown in Fig.
\ref{rhoaxt}. Unlike the contact process, in which $\rho$ follows
a simple power law before attaining the QS regime, in the SRW model
the relaxation is more complicated.  Absence of simple power-law relaxation
of the activity density has also been noted for stochastic sandpiles \cite{cssp1d}.
We calculate the local slope $\delta(t)$ via
piece-wise linear fits to the data for $\ln \rho$ versus $\ln t$
on the interval $[t/\alpha, \alpha t]$, with $\alpha = 1.76$
(that is, about 30 data points, with an increment of 0.1 in $\ln t$).
The inset of Fig.~\ref{rhoaxt} shows $\delta(t)$ decreasing
systematically with time, from around $0.145$ to approximately
$0.12$ over the interval studied. (Note that the data for longer
times come from larger systems.)
If we associate with the short- and long-time regimes the values
$\delta_{short} \approx 0.143(3)$ and $\delta_{long} \approx
0.121(3)$, then the scaling relation
$\delta=\beta/\nu_\parallel=\beta/(\nu_\perp z)$ yields
$\delta=0.212/1.50=0.14$, in good agreement with the short-time
value. Plotting $\rho^* = L^{\beta/\nu_\perp} \rho$ as a function of
$t^* = t/L^z$ we observe a good collapse of the data for different
system sizes in the first regime using $z=1.50$, and in the second
regime if we use $z=1.72$. In both cases the best collapse is
obtained using $\beta/\nu_\perp=0.215$.

\begin{figure}[!h]
\includegraphics[width=10cm]{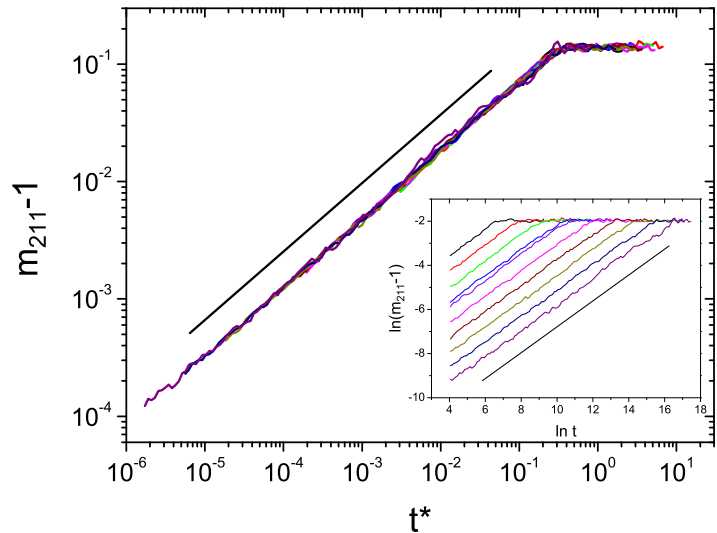}
\caption [cruzm]{Scaling plot of
$m_{211}-1$ versus $t^*=t/L^z$ at the critical point, using
$z=1.70$, for system sizes up to $32\,000$. The inset shows the
unscaled data for $100\leq L \leq 32000$ (from top to bottom). The
slope of the straight lines is
$0.587$. } \label{m211xtnovo}
\end{figure}

In contrast to the complicated behavior of the activity density,
the quantity $m_{211} (t) -1$ exhibits {\it simple} power-law scaling over
four or more decades. Using $t^*=t/L^z$ with $z=1.71$, we obtain a
good data collapse for all system sizes studied, as shown in
Fig. \ref{m211xtnovo}.  The relation $m_{211}-1\sim t^{1/z}$ yields
$z = 1.70(1)$, consistent with the collapse values for $m_{211}$ and $\rho(t)$
in the long-time regime, but clearly inconsistent with $z=1.50(4)$ found using FSS
analysis in the QS regime. Curiously, $m_{211}$ shows no hint of the
crossover exhibited by the order parameter.  Note that the value
$\delta=0.121(2)$ associated with the second regime of the order
parameter is consistent with $\beta/\nu_\perp = 0.212$ and $z=1.71$.
Thus the scaling of $m_{211}$ follows, from the beginning, that
observed in $\rho$ in the long-time regime.

\subsection{Comparison of exact QS and simulation results}

In Table~\ref{compQSsim} we compare results obtained via
exact analysis of small systems (QSA) and simulation.
The results are consistent to within
uncertainty {\it except} for the dynamic exponent $z$.
The QSA predictions for $\beta/\nu_\perp$ and
(especially) $z$ seem less reliable than those derived from
simulation.  On the other hand, the QSA estimates for moment rations
are nominally of higher precision than the simulation results.
Care must be exercised, however, since the QSA analysis may
be subject to relatively large
finite-size corrections.  (The present study and previous works on
models in the CDP class suggest that corrections to scaling and
finite-size effects are stronger for this class than for the contact
process.)  Table~\ref{compQSsim} also includes results
on sandpiles and the conserved lattice gas.  The agreement between these
studies and the present work is quite good, leaving little doubt that the SRW
model belongs to the same universality class as stochastic conserved
sandpiles and conserved directed percolation.

\begin{table}[h]
\centering \caption[tabela]{\footnotesize Comparison of critical properties of the SRW
model found via exact analysis of small systems (QSA) and Monte Carlo simulation (MC),
and results from previous studies (Prev) on models in the CDP class: $a$: restricted
stochastic sandpile \cite{sandpile2006}; $b$: conserved lattice gas \cite{kockelkoren};
$c$: restricted stochastic sandpile \cite{unpub}.
}
\label{compQSsim}
\begin{center}
\begin{tabular}{|c|c|c|c|} \hline
Quantity           &     QSA     &    MC        &   Prev \\ \hline\hline
$\lambda_c$        & 0.09002(10) & 0.090085(12) &              \\
$\beta/\nu_\perp$  & 0.241(1)    & 0.212(6)     & 0.213(6) $^a$ \\
$\beta$            &             & 0.290(4)     & 0.289(12)$^a$ \\
$\gamma/\nu_\perp$ & 0.541(2)    & 0.58(1)      & 0.55(1) $^c$ \\
$\nu_\perp$        & 1.28(1)     & 1.33(5)      & 1.36(2)  $^a$  \\
      $z$          & 1.664(4)    & 1.50(4)      & 1.55(3)  $^b$  \\
$m_{211}$          & 1.142(1)    & 1.141(8)     & 1.142(8) $^a$ \\
$m_{3111}$         & 1.420(5)    & 1.415(26)    & 1.425(25)$^c$  \\
$m_{-1}m$          & 1.308(12)   & 1.327(27)    & 1.332(10)$^c$  \\
$q_c$              & -0.454(2)   & $-0.47(3)$   & -0.46(3) $^c$  \\ \hline
\end{tabular}
\end{center}
\end{table}

\section{Discussion}

We study sleepy random walkers (SRW) in one dimension
using (numerically) exact quasistationary analysis of small systems
and Monte Carlo simulation.
Based on considerations of symmetry and conserved quantities, one expects
the SRW process to belong to the conserved directed percolation (CDP) class,
typified by conserved stochastic sandpiles.
Our results for critical exponents and
moment ratios support this conclusion.
Different from most examples of the CDP class studied until now, the SRW
process includes a continuously-variable control parameter which facilitates
simulation and numerical analysis.

The present work represents a further test of the exact QS analysis
proposed in \cite{exact sol}.
In addition to locating the critical
point with good precision, QS analysis furnishes
fair results for the critical exponent
$\nu_\perp$ and quite good estimates for the moment ratios $m_{211}$ and $m_{3111}$.
While somewhat better than the preliminary study of a model in the CDP class, QSA
predictions for the SRW model are not of the quality obtained for the contact
process \cite{exact sol}.  This appears to be connected with the stronger finite-size effects
and corrections to scaling observed for models in the CDP class.
Exact analysis of QS properties is nevertheless a useful
complement to simulation, as in this method the long-time behavior
(conditioned on survival) is surely attained, whereas simulations
are subject to the nagging possibility of insufficient relaxation
time.  The QS calculations typically require (for the largest system studied) several days on a
reasonably fast computer, that is, a small fraction of the time
invested in a simulation study.

In Monte Carlo simulations, we test various scaling relations via data collapse,
in both the sub- and supercritical
regimes, and compare the resulting critical exponents with those obtained
via finite-size scaling at the critical point.
The results are generally consistent between the regimes, as well as with those
of previous studies of stochastic conserved sandpiles.
Despite the general agreement, we identify several inconsistencies and examples
of anomalous behavior.  First, the estimates for the exponent $\nu_\perp$ in the
inactive phase are significantly smaller (by roughly 5\%) than those obtained at the
critical point or in the active phase.
Second, and more significantly,
the relaxation of the order parameter to its quasistationary
value at the critical point is marked by two apparent scaling
regimes, with associated exponents
$z_{short} = 1.50$, $\delta_{short} = 0.143$, $z_{long} = 1.71$, and
$\delta_{long} = 0.121$, respectively. Paradoxically, $z_{short}$ is
close to the dynamic exponent characterizing finite-size scaling in
the asymptotic long-time (i.e., quasistationary) regime. A similar
two-regime relaxation is observed in the one-dimensional conserved
stochastic sandpile \cite{cssp1d}; the latter study reports
$z_{long} = 1.75(3)$ and somewhat larger values for the exponents
$\delta$.  Despite these minor numerical differences it seems likely
that anomalous relaxation is a characteristic of the CDP class in general.
Finally, we have noted a possible violation of scaling associated with
the scaled variance of the order parameter $\chi$ in the inactive phase.

Given the complex pattern of relaxation of the order parameter, it is
surprising that $m_{211}-1$, another quantity expected to exhibit
power-law scaling at the critical point, in fact shows a
near-perfect data collapse in a {\it single} scaling regime that
corresponds essentially to the two scaling regimes of the order parameter.
The associated dynamic exponent is $z=1.71(1)$, the same as $z_{long}$ to within uncertainty.
This exponent is however considerably larger than the dynamic exponent $z=1.50(4)$
associated with finite-size scaling at the critical point.
As discussed in \cite{scaling}, the
difference may reflect the existence of two time scales, one associated with the
relaxation of the order parameter to its QS value, the other
related to the finite-size lifetime.
These two times scale in the same manner at absorbing phase transitions
in models without a conserved density, such as the CP.
These unexpected findings should motivate further study of the SRW and related models,
with the goal of a complete and coherent picture of scaling in the CDP
universality class.  From the present vantage it appears that such a
picture will be significantly more complex than for directed
percolation.
\vspace{1cm}

\noindent {\bf Acknowledgment}

We are grateful to CNPq, Brazil, for financial support.


\end{document}